# Ultra-wideband Antireflection Assisted by Continuously Varying Temporal Medium

Yi Zhang[1], Liang Peng[2,3], Zhengjie Huang[1], Lixin Ran[1] and Dexin Ye[1,4]

[1] *Laboratory of Applied Research on Electromagnetics, Zhejiang University, Hangzhou 310027, China*
[2] *School of Information and Electrical Engineering, Zhejiang University City College, Hangzhou 310015, China*
[3] *pengl@zucc.edu.cn*
[4] *desy@zju.edu.cn*

**Abstract**: We demonstrate that reflectionless propagation of electromagnetic waves between two different materials can be achieved by designing an intermediate temporal medium, which can work in an ultra-wide frequency band. Such a temporal medium is designed with consideration of a multi-stage variation of the material's permittivity in the time domain. The multi-stage temporal permittivity is formed by a cascaded quarter-wave temporal coating, which is an extension of the antireflection temporal coating by Pacheco-Peña *et al* [[1] Optica 7, 323 (2020)]. The strategy to render ultra-wideband antireflection temporal medium is discussed analytically and verified numerically. In-depth analysis shows that the multi-stage design of the temporal media implies a continuously temporal variation of the material's constitutive parameters, thus an ultra-wideband antireflection temporal medium is reasonably obtained. As an illustrative example for application, the proposed temporal medium is adopted to realize impedance matching between a dielectric slab and free space, which validates our new findings.

## 1. Introduction

An electromagnetic (EM) wave is commonly reflected when it impinges on an interface in-between two different media [2]. Since reflection used to bring in some undesired effects that perturb the EM system, it has been an interesting and important topic to realize reflectionless transmission in a wide frequency band, as usually requested in the EM engineering. The typical and simplest approach to solve this issue could be the utilization of antireflection coating, e.g., the impedance transformer by a quarter-wave (QW) slab [3], with which EM waves reflected by its front and back surface cancel each other due to the destructive interference. Nevertheless, the application of the QW slab is limited in practice because it works well only for specific frequencies, as well as the predefined incident angle. Although much effort has been made to overcome the drawbacks of the QW slab, such as employing multiple layers [4], varying refractive index profiles [5, 6], and changing the surface textures [7], none of the approaches are satisfactory in considering both the operation spectral band and the dynamic range of incident angle. Remarkably, in 2018, K. Im *et al*. proposed a universal antireflection layer design by utilizing the spatially dispersive media, which implies a possible (almost perfect) method to render reflectionless propagation of EM waves regardless of the polarization and angle of incidence [8]. However, to fabricate such a spatially dispersive layer is rather challenging, due to the material complexity.

Recently, the time-varying media, whose constitutive parameters vary with time, are proposed and have been attracting lots of attentions [9]. Prior to the conventional media existing naturally (which is time-invariant), the powerful time-varying media have been suggested in various applications to boost the degree of EM wave manipulation, such as temporally-modulated surface [10] and metamaterials [11], energy accumulation [12, 13], time reversal operation [14], inverse prism [15, 16], magnetless nonreciprocity [17-19], wave pattern engineering [20], temporal photonic crystal [21, 22], chiral reflection [23], impedance matching [8], and designing higher-order transfer functions [24]. In particular, V. Pacheco-Peña *et al*. proposed a temporal analogue of the QW impedance transformer, indicating an alternative approach to achieve reflectionless penetration of EM waves under arbitrary incident angles [1]. The originally proposed temporal coating relies on the destructive interference between two backward waves, and hence is also restricted to work in a narrow band close to the frequency. It would be an interesting and open question that how such a peculiar temporal operation can be extended to render wideband reflectionless propagation, as naturally requested by the practical engineering.

In this paper, we theoretically demonstrate that by continuously changing its dielectric constant, a temporal material can boost the nearly perfect EM wave penetration between two different media in an ultra-wide frequency band. Theoretically, this time-varying media is derived from an elaborately designed multi-stage (discrete) temporal operation, which extends the concept of QW transformer. It is shown that the working band of the multi-stage time-varying media is enlarged with increasing the number of temporal QW stages. At the very limit, a continuously variant temporal operation provides an ultra-wide operation band for reflectionless wave propagation. Theoretical analyses coincide with our numerical and full-wave simulations, verifying our new findings.

## 2. Theory and design

Without loss of generality, we begin with the concept of a time-varying medium, where the time-dependent permittivity of a spatially unbounded medium changes suddenly everywhere at a particular time. As shown in the top panel of Fig. 1(a), the initial relative permittivity of a nonmagnetic medium is $\varepsilon_0$ and changes sharply to $\varepsilon_f$ at $t = t_b$. With a monochromatic wave travelling in such a medium, the temporal boundary leads to a forward wave (FW) and a backward wave (BW) due to the different wave impedances before and after the temporal boundary [25]. Essentially, it arises from the conditions that the electric displacement ***D*** and magnetic induction ***B*** are continuous across the temporal boundary [26, 27]. Consequently, the frequency of the travelling monochromatic wave is changed to $\omega_f = (\varepsilon_0/\varepsilon_f)^{1/2}\omega_0$ away from $\omega_0$. Although the frequency is changed, the FW and BW can be effectively considered as the reflected and transmitted waves at the temporal boundary, respectively. Defined by the ratios of the electric field amplitudes of the FW and BW to that of the initially travelling wave [26, 28], the "temporal Fresnel" coefficients can be described as

$$R = (\varepsilon_0/\varepsilon_f - \sqrt{\varepsilon_0/\varepsilon_f})/2,$$
$$T = (\varepsilon_0/\varepsilon_f + \sqrt{\varepsilon_0/\varepsilon_f})/2$$
(1)

where R and T denote the temporal reflection and transmission coefficients, respectively.

As proposed in ref. [1], the BW induced by the temporal boundary can be eliminated by introducing a QW intermediate stage with a relative permittivity $\varepsilon_{inter} = (\varepsilon_0\varepsilon_f)^{1/2}$ and a time duration $\tau_{inter} = T_{inter}/4 = (\varepsilon_{inter}/\varepsilon_0)^{1/2}/4f$. Here, $T_{inter}$ denotes the period of the travelling wave in the intermediate stage. As shown in the panel-(I) of Fig. 1(b), the intermediate stage splits the single temporal boundary into two ones, and the initially travelling EM wave will be split into two BWs and two FWs, respectively. The phase delay between two temporal boundaries is denoted as $\phi_{inter} = \omega_0(\varepsilon_0/\varepsilon_{inter})^{1/2}\tau_{inter} = \pi/2$. By using equation (1), the total temporal reflection coefficient $R_{total}$ and transmission coefficient $T_{total}$ can be derived as

$$R_{total} = T_1 e^{j\phi_{inter}} R_2 + R_1 e^{-j\phi_{inter}} T_2 = T_1 e^{j\pi/2} R_2 + R_1 e^{-j\pi/2} T_2$$
$$T_{total} = T_1 e^{j\phi_{inter}} T_2 + R_1 e^{-j\phi_{inter}} R_2 = T_1 e^{j\pi/2} T_2 + R_1 e^{-j\pi/2} R_2$$
(2)

Here, $R_n$ and $T_n$ are the temporal transmission and reflection coefficients on the *n*-th temporal boundary. As $R_1 = R_2$ and $T_1 = T_2$, two BWs have the same amplitudes but π-phase difference, and thus cancel each other completely. In this case, the intermediate layer behaves as the temporal QW impedance transformer, which works in some specified frequencies [1].

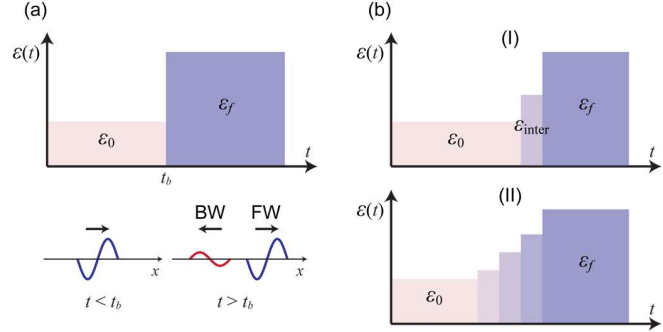

Fig 1. (a) A temporal boundary where the time-dependent permittivity is changed from $\varepsilon_0$ to $\varepsilon_f$ at $t = t_b$, produces a forward wave (FW) and a backward wave (BW) due to the impedance mismatching. (b) Panel-(I) shows the single quarter-wave (QW) temporal impedance transformer scheme, and panel-(II) shows the multi-stage one for wideband impedance matching.

In the spatial counterparts, although a single QW impedance transformer works only in a narrow band, broadband impedance matching may be rendered by a cascade of multi-stage QW transformers. Inspired by this, we consider designing a temporal transformer through multi-stage temporal QW transformers, to seek the solution for broadband impedance matching.

Let's assume there exists an $N$-order transformer, which consists of $N$ intermediate stages together with $N+1$ temporal boundaries, as shown in the panel-(II) of Fig. 1(b). The relative permittivity of the $n$-th stage is $\varepsilon_n$ with a time duration $\tau_n$ ($n = 1, 2, 3, …, N$), and thus the corresponding wave impedance is $Z_n = (\varepsilon_0/\varepsilon_n)^{1/2} Z_0$. Here, $Z_0$ is the wave impedance in the initial stage. The phase delay between the $n$-th and $(n+1)$-th boundaries is represented by $\phi_n = \omega_0(\varepsilon_0/\varepsilon_n)^{1/2}\tau_n$. Since every component of the travelling waves will be split into two components when passing through a temporal boundary, there will be $2^{N+1}$ components in total, and half of them are BWs. In such a case, the total reflection and transmission coefficients can be derived based on the recursive relation

$$R_{total}^n = T_{total}^{n-1} e^{j\phi_{n-1}} R_n + R_{total}^{n-1} e^{-j\phi_{n-1}} T_n,$$
$$T_{total}^n = T_{total}^{n-1} e^{j\phi_{n-1}} T_n + R_{total}^{n-1} e^{-j\phi_{n-1}} R_n, \quad (3)$$

where $R^n_{total}$ and $T^n_{total}$ denote the total reflection and transmission coefficients when the EM wave passes through first $n$ boundaries. $R_n = [\varepsilon_{n-1}/\varepsilon_n - (\varepsilon_{n-1}/\varepsilon_n)^{1/2}]/2$ and $T_n = [\varepsilon_{n-1}/\varepsilon_n + (\varepsilon_{n-1}/\varepsilon_n)^{1/2}]/2$ are the temporal reflection and transmission coefficients on the $n$-th boundary, respectively.

Suppose that the difference between two adjacent stages is small, we can safely find out the derivatives of $R^n_{total}$ and $T^n_{total}$, and all the quantities can be approximately obtained based on the small reflection theory [3]. Within the first order precision, i.e., retaining the product terms with only one reflection coefficient $R_n$, the total reflection coefficient $R^N_{total}$ can be approximated by

$$R_{total}^N \approx (\prod_{n=1}^{N+1} T_n) e^{-j\sum_{n=1}^{N}\phi_n} \cdot (\Gamma_0 + \Gamma_1 e^{j2\phi_1} + \Gamma_2 e^{j2(\phi_1+\phi_2)} + ... + \Gamma_N e^{j\sum_{n=1}^{N}2\phi_n}), \quad (4)$$

where $\Gamma_n = R_{n+1} / T_{n+1} = (Z_{n+1} - Z_n)/(Z_{n+1} + Z_n)$. Conceptually, the spectral distribution of $R^N_{total}$ may be optimized by tuning the permittivity $\varepsilon_n$ and the duration time $\tau_n$ for each stage, and hence an ultra-wideband feature could be rendered. It is quite interesting that the particular expansion of $R^N_{total}$ is rather helpful in seeking the optimal solution of $\varepsilon_n$ and $\tau_n$.

We can see that $R^N_{total}$ takes the form like the $N$-order binomial function

$$\gamma(\phi) = A(1 + e^{j2\phi})^N = A\sum_{n=0}^{N} C_N^n e^{j2n\phi}, \quad (5)$$

where $A$ is a constant. Reflectance satisfying this expansion is always zero at $\phi = \pi/2$ and exhibits flat close to this point since the 1st order derivative of $|\gamma(\phi)|$ is tiny [3]. Apparently, expression (5) represents a passband filter effect of the reflectance, and higher $N$ means a wider passband. As a direct illustration, here we draw the magnitudes of binomial functions with different orders ($N$ = 1, 5, 9 and 17) in Fig. 2(a). Obviously, the bandwidth can be tuned by varying $N$.

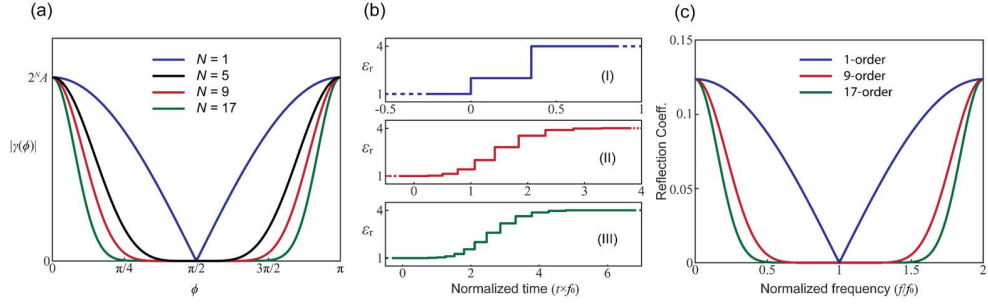

Fig. 2. (a) The magnitudes of binomial functions $|\gamma(\phi)|$ for different orders ($N$ = 1, 5, 9 and 17). (b) The calculated permittivity profiles of 1-order [Panel-(I)], 9-order [Panel-(II)], and 17-order [Panel-(III)] multi-stage temporal transformers according to equations (6). (c) The calculated reflection coefficient magnitudes versus frequency for different multi-stage temporal transformers according to equation (4).

Previous analysis shows that a broadband reflectionless propagation of EM waves would be possible, provided that $R^N_{total}$ takes the form identical to equation (5), which indicates a multi-stage temporal medium. By comparing equations (4) and (5), we get $\phi_n = \omega_0(\varepsilon_0/\varepsilon_n)^{1/2}\tau_n = \pi/2$ and $\Gamma_n = (Z_{n+1} - Z_n) / (Z_{n+1} + Z_n) = AC_N^n$, or equivalently $Z_{n+1}/Z_n = (1+AC_N^n) / (1-AC_N^n)$. Here, the constant $A$ can be numerically determined based on the bisection method, with considering $Z_f/Z_0 = \Pi_{n=0}^N (1+ AC_N^n) / (1- AC_N^n)$. For a given operating frequency $\omega_0$, the permittivity $\varepsilon_n$ and duration time $\tau_n$ of each stage can be safely extracted by

$$\begin{cases} Z_n = Z_{n-1} \times (1+ AC_N^n)/(1- AC_N^n) \\ \varepsilon_n = \varepsilon_0 Z_0^2 / Z_n^2 \\ \tau_n = \sqrt{\varepsilon_n}\, \pi/(2\omega_0 \sqrt{\varepsilon_0}) \end{cases} \qquad (6)$$

In the numerical evaluations, a series of multi-stage temporal binomial transformers with $N$ = 1, 9 and 17 are designed by obeying equation (6). Let $\varepsilon_0$ = 1, $\varepsilon_f$ = 4, and the center operating frequency $f_0$, the requested permittivities $\varepsilon_n$ and duration times $\tau_n$ distributions are listed in Table. 1, as plotted in Fig. 2(b). With equation (4), the reflection coefficient is predicted as shown in Fig. 2(c). It is clear that the 1-order transformer is identical to a single QW transformer [1], and its relative bandwidth for the reflection below -40 dB ($|R_{total}| < 0.01$) is only 10.2%. For the 9-order and 17-order transformers, the relative bandwidths can reach up to 112% and 132%, showing an enhanced broadband antireflection feature.

Apparently, as the order number $N$ increases, the step changes in permittivity between two adjacent stages become smaller, and the multi-stage discrete transformer approaches to a continuous one with an infinitely long duration time. Although a continuous variation of the temporal medium is possible, the infinite time duration is unreachable in practice. It would be quite helpful if a broadband temporal medium could be rendered within a finite time duration.

Table 1. Relative permittivities and duration times of 1-order, 9-order, 17-order transformers

| | | |
|---|---|---|
| $\varepsilon_{rn}$ 1-order | 2 | |
| $\tau_n(1/f_0)$ | 0.3536 | |

| $\varepsilon_{rn}$ 9-order | 1.0027 | 1.0274 | 1.1324 | 1.4215 | 2 | 2.8140 | 3.5322 | 3.8932 | 3.9892 |
|---|---|---|---|---|---|---|---|---|---|
| $\tau_n(1/f_0)$ | 0.2503 | 0.2534 | 0.2660 | 0.2981 | 0.3536 | 0.4194 | 0.4699 | 0.4933 | 0.4993 |
| $\varepsilon_{rn}$ 17-order | 1.0000 | 1.0002 | 1.0016 | 1.0089 | 1.0345 | 1.1045 | 1.2588 | 1.5463 | 2 |
| $\tau_n(1/f_0)$ | 0.2500 | 0.2500 | 0.2502 | 0.2511 | 0.2543 | 0.2627 | 0.2805 | 0.3109 | 0.3536 |
| $\varepsilon_{rn}$ 17-order | 2.5867 | 3.1775 | 3.6217 | 3.8664 | 3.9649 | 3.9935 | 3.9992 | 4.0000 | |
| $\tau_n(1/f_0)$ | 0.4021 | 0.4456 | 0.4758 | 0.4916 | 0.4978 | 0.4996 | 0.5000 | 0.5000 | |

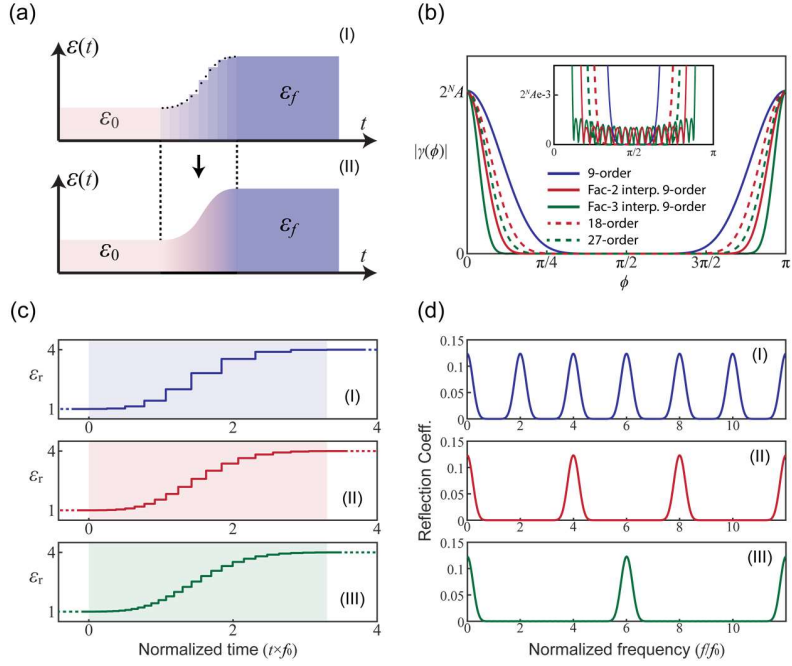

Fig. 3. (a) The interpolation approach for rendering a continuous temporal transformer. (b) The magnitudes of binomial functions $|\gamma(\phi)|$ with different interpolation factors. (c) The calculated permittivity profiles of Panel-(I): 9-order temporal transformer. Panel-(II): Factor-2 interpolated 9-order temporal transformer. Panel-(III): Factor-3 interpolated 9-order temporal transformer. (d) The reflection coefficients of transformers in (c).

Therefore, we use an interpolation-like method to smooth the temporal permittivity profile and expand the bandwidth while keeping the total duration time unchanged, as illustrated in Fig. 3(a). The key concept is to insert infinite items with general binomial coefficients for rational value in the previous $N$-order binomial function. The factor-$(1/\Delta n)$ interpolated $N$-order binomial function is then taking the form of

$$\gamma(\phi) = A\sum_{l=0}^{N/\Delta n} C_N^{l\Delta n} e^{j2l\phi} , \tag{7}$$

where $1/\Delta n$ is an integer and denotes the interpolation factor. Obviously, with the interpolation factor $1/\Delta n = 1$, expression (7) agrees with equation (5). To illustrate the difference before and after interpolation, we show the amplitude of $\gamma(\phi)$ with different orders as well as different interpolation factors in Fig. 3(b). As expected, the relative bandwidth can be broadened by either increasing the order number $N$ or the interpolation factor $1/\Delta n$. However, compared with the 18-order transformer, the factor-2 interpolated 9-order one provides a wider relative bandwidth, and so it is for the factor-3 interpolated 9-order transformer compared with the 27-order one. For the interpolated transformer, the trade-off is the emergence of inevitable nonzero ripples in the passband. It should be emphasized that the magnitudes of reflection ripples depend on the order number $N$.

Similarly, by replacing $n$ with $l\Delta n$ in equations (6), an interpolated multi-stage transformer can be obtained. Here we design the 9-order transformers with different interpolation factors ($1/\Delta n$ = 1, 2 and 3), and keep the total duration time constant ($\tau_{total} = 3.303/f_0$). The derived permittivity profiles and the corresponding reflection coefficient magnitudes are shown in Fig. 3(c) and 3(d). As illustrated, the increasing of the interpolation factor is equivalent to promoting the center frequency ($\omega_0$), but has almost no influence on the reflectance in the lower frequency band. So, the antireflection is effectively extended to high frequencies. It can also be imagined that, when the interpolation becomes subtle, the multi-stage temporal medium is asymptotic to a continuous temporal transformer, which is able to provide a wide passband as expected.

## 3. Full wave simulation

To verify our previous analysis, full-wave simulations are performed based on the commercial software COMSOL® Multiphysics. The initial travelling Gaussian pulse electric field takes the form of $E_y(x, t) = \exp[-(x-ct)^2/(2\sigma_x^2)]\exp[-jk_0(x-ct)]$, and thus its spectrum is $E_y(x, f) = (2\pi)^{1/2}\sigma_x/c \cdot \exp[-x^2/(2\sigma_x^2) - jk_0x] \cdot \exp\{-4\pi^2\sigma_x^2/(2c^2) \cdot [f - (\sigma_x^2 k_0 c - jxc)/(2\pi\sigma_x^2)]^2\}$, with a center frequency $f_c = k_0c/2\pi$. In addition, from the second exponential term of $E_y(x, f)$, the spectral deviation of such pulse can be calculated as $\sigma_f = c/(2\pi\sigma_x)$.

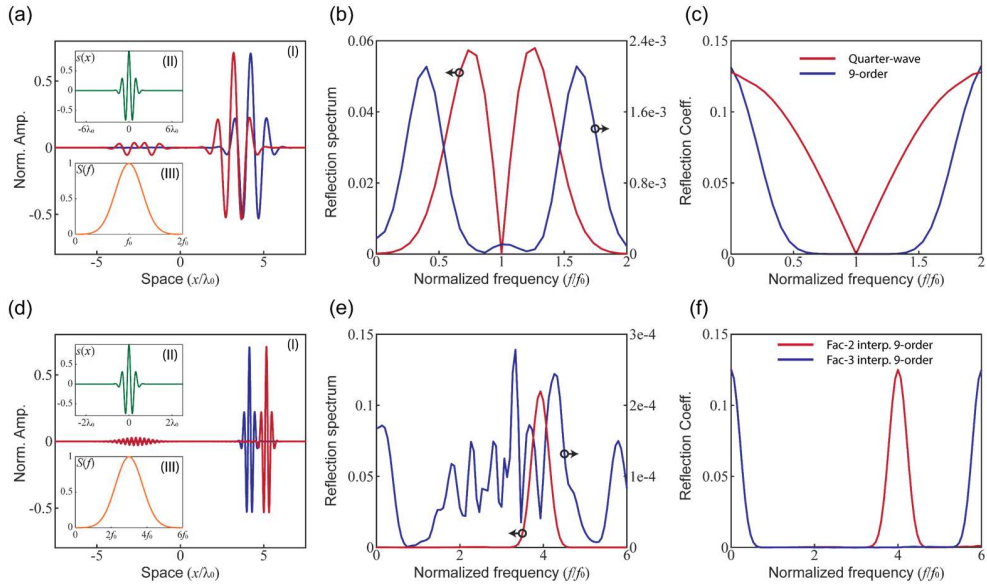

Fig. 4. Full-wave simulations. (a) Panel-(I): The electric field distributions of a Gaussian pulse passing through the quarter-wave temporal transformer (red line) and the 9-order multi-stage transformer (blue line). Panel-(II): The waveform of the testing Gaussian pulse centered at $f_0$ with a spectral deviation of $0.25f_0$. Panel-(III): The spectrum of the Gaussian pulse in (a)-(II). (b) The spectra of reflection in (a). (c) The calculated reflection coefficients from (b) and (a)-(III). (d) Panel-(I): The electric field distributions of a Gaussian pulse passing through the factor-2 interpolated 9-order transformer (red line) and factor-3 interpolated 9-order transformer (blue line). Panel-(II): The waveform of the testing Gaussian pulse centered at $3f_0$ with a spectral deviation of $0.75f_0$. Panel-(III): The spectrum of the Gaussian pulse in (d)-(II). (e) The spectra of reflection in (d). (f) The calculated reflection coefficients from (e) and (d)-(III).

First, we set the center frequency to be $f_0$ with a spectral deviation $\sigma_f = c/(2\pi\sigma_x) = 0.25f_0$ to characterize the responses of previous 1-order and 9-order transformers, respectively. Panels (I) and (II) of Fig. 4(a) show the spatial distribution of such a Gaussian pulse electric field at $t = 0$, and its spectrum (a function of the normalized frequency $f/f_0$). In the simulation, $\varepsilon_0 = 1$ and

$\varepsilon_f = 4$ are set. The time-dependent permittivities $\varepsilon(t)$ for the transformers are chosen with $N = 1$ and $N = 9$, as shown in Figs. 2(a) and 2(b). The red (blue) line in Fig. 4(a) shows the spatial electric field distributions when the Gaussian pulse passes through the $N = 1(9)$ transformer at $t = 6T_0$ ($7.5T_0$). As we imagine, obvious BW is observed for the 1-order transformer due to its narrow passband width, while there is no obvious BW observed for the $N = 9$ case. For a better interpretation, we plot the calculated spectral contents of two BWs and the corresponding temporal reflection coefficients in Fig. 4(b) and 4(c), respectively. The calculated temporal reflection coefficients are agreeing very well with those shown in Fig. 2(c).

Second, another set of simulations is performed to characterize the responses of $N = 9$ temporal transformers with different interpolation factors ($1/\Delta n = 2, 3$). The center frequency is set to be $f_c = kc/2\pi = 3f_0$, and the spectral deviation $\sigma_f = c/(2\pi\sigma_c) = 0.75f_0$. Similar to Fig. 4(a), panels (I) and (II) of Fig. 4(d) show the spatial distribution at $t = 0$ and its spectrum. The time-dependent permittivities $\varepsilon(t)$ for the factor-2 and factor-3 interpolated 9-order transformer are the same as those shown in panels (II) and (III) of Fig. 3(c), respectively. The red line and the blue line in Fig. 4(d) show the spatial electric field distributions at $t = 9T_0$ for the factor-2 interpolated 9-order transformer, and $t = 7T_0$ for the factor-3 interpolated one. The spectrum of two BWs and the temporal reflection coefficients are shown in Fig. 4(e) and 4(f). It is evident that these simulated results agree well with those calculated based on the small reflection theory.

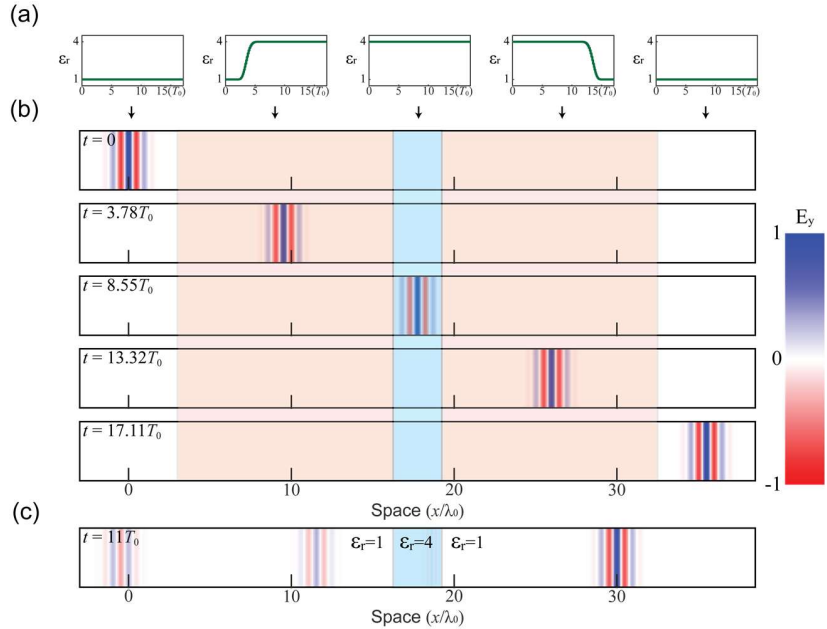

Fig. 5. The temporal matching of a dielectric slab (a) The permittivity profiles of constant air ($\varepsilon_r = 1$, white regions), temporal matching media (time-varying $\varepsilon_r$, orange regions) and dielectric slab ($\varepsilon_r = 4$, blue regions). (b) The electric field spatial distribution at $t = 0$, $3.78T_0$, $8.55T_0$, $13.32T_0$ and $17.11T_0$, respectively, in the presence of temporal matching media. (c) The electric field spatial distribution at $t = 11T_0$ in the absence of temporal matching media, in which multiple reflections can be observed.

Finally, we present a typical example that the proposed antireflection time-varying media can achieve antireflection propagation of a pulse signal (with ultrawide spectrum) illuminating on a dielectric slab. As shown in Fig. 5(a). The dielectric slab (blue region) occupies the interval of $16.25 < x/\lambda_0 < 19.25$, whose relative permittivity is $\varepsilon_r = 4$. Two temporal transformers (orange regions) are placed symmetrically on both sides of the dielectric slab to eliminate reflection, i.e., in regions $3 < x/\lambda_0 < 16.25$, and $19.25 < x/\lambda_0 < 32.5$. The white regions denote the free space. The incident Gaussian pulse (electric field intensity) identical to the one shown in the panel-(II) of Fig. 4(d), travels from the left to the right. The time-dependent permittivity $\varepsilon(t)$ for each region is shown in Fig. 5(a). The left side temporal transformer possesses $\varepsilon(t)$ identical to those shown in the panel-(II) of Fig. 3(c), with starting at $t = 2T_0$ and ending at $t = 5.55T_0$. This setting guarantees that the permittivity begins to change only if the whole Gaussian pulse has entered the temporal matching medium on the left side. Meanwhile, the permittivity profile of the right temporal transformer is a reversion of the left one, starting at $t = 11.55T_0$ and ending at t = $15.1T_0$. With this setting, we ensure that the permittivity begins to change when the Gaussian pulse has entered the transformer on the right side completely.

Fig. 5(b) shows the simulated electric field distributions at different times (t = 0, 3.78, 8.55, 13.32, and 17.11 $T_0$). As we expected, a negligible reflected wave can be seen when two antireflection time-varying media are used. For comparison, we also show the electric field distribution at t = $11T_0$ while removing two antireflection time-varying media. In such a case, several reflected pulses and transmitted pulses have been observed, arising from the multiple reflections between two spatial boundaries. It should be noted that the amplitude of the transmitted wave in the dielectric slab is reduced, due to the conservation of electric displacement [26, 28]. Consequently, since the permittivity of the right antireflection temporal medium is changed from a higher one to a smaller one, the final amplitude of the transmitted wave recovers to unity, achieving a broadband full transmission without frequency conversion.

## 4. Conclusion

In conclusion, we demonstrate the design of multi-stage and/or continuous temporal transformers, which can be used to realize antireflection propagation of EM waves in an ultra-wideband. The proposed transformer is the direct extension of the standard temporal QW impedance transformer, which guarantees the impedance matching nature at the reference frequencies. We discuss the mechanism of the broadband operation, and the manipulation of the spectral responses of the transformer. It is shown that the proposed temporal transformer is effective to EM pulses with arbitrary waveforms, provided that the signal spectrum is limited to the allowed range (passband). Compared with the multi-stage transformer, the continuously varying temporal transformer has a consecutive (rather than periodically distributed) passband in higher frequency, which is rather attractive and meaningful for practical applications. Conceptually, the continuous temporal transformer is a good candidate for achieving the spatiotemporal impedance matching functionality. The temporal transformer discussed in this paper, multi-stage and/or continuous, shows compact and effectiveness in realizing antireflection delivery of EM signals between two different systems, which could be potentially applied to various scenarios, including the matching ending, power collection/detection, EM wave transformer, and the EM isolators, etc.

## Funding

This work is supported by the National Natural Science Foundation of China (NSFC) under grants 62071420, 62071417, 62122068, and 61875051, and the Natural Science Foundation of Zhejiang Province (ZJNSF) under grant LR21F010002.

## Acknowledgments

Y. Z., L. P., D. Y. designed the research. All authors contributed to data interpretation and the composition of the manuscript.

## Disclosures

The authors declare no conflicts of interest.

## References

1. V. Pacheco-Peña and N. Engheta, "Antireflection temporal coatings," Optica 7, 323-331 (2020).
2. L. D. Landau, The classical theory of fields (Elsevier, 2013), Vol. 2.
3. D. M. Pozar, Microwave engineering (John wiley & sons, 2011).

4. H. A. Macleod and H. A. Macleod, Thin-film optical filters (CRC press, 2010).
5. Q. Tang, S. Ogura, M. Yamasaki, and K. Kikuchi, "Experimental study on intermediate and gradient index dielectric thin films by a novel reactive sputtering method," Journal of Vacuum Science & Technology A: Vacuum, Surfaces, and Films 15, 2670-2672 (1997).
6. H. K. Raut, V. A. Ganesh, A. S. Nair, and S. Ramakrishna, "Anti-reflective coatings: A critical, in-depth review," Energy & Environmental Science 4, 3779-3804 (2011).
7. P. Spinelli, M. Verschuuren, and A. Polman, "Broadband omnidirectional antireflection coating based on subwavelength surface Mie resonators," Nature communications 3, 1-5 (2012).
8. K. Im, J.-H. Kang, and Q.-H. Park, "Universal impedance matching and the perfect transmission of white light," Nature Photonics 12, 143-149 (2018).
9. E. Galiffi, R. Tirole, S. Yin, H. Li, S. Vezzoli, P. A. Huidobro, M. G. Silveirinha, R. Sapienza, A. Alù, and J. Pendry, "Photonics of time-varying media," Advanced Photonics 4, 014002 (2022).
10. D. Oue, K. Ding, and J. B. Pendry, "Calculating spatiotemporally modulated surfaces: A dynamical differential formalism," Physical Review A 104, 013509 (2021).
11. P. A. Huidobro, M. G. Silveirinha, E. Galiffi, and J. B. Pendry, "Homogenization Theory of Space-Time Metamaterials," Physical Review Applied 16, 014044 (2021).
12. J. Pendry, E. Galiffi, and P. Huidobro, "Gain in time dependent media - a new mechanism," Journal of the Optical Society of America B 38, 3360-3366 (2021).
13. K. A. Lurie and V. V. Yakovlev, "Energy accumulation in waves propagating in space-and time-varying transmission lines," IEEE Antennas and Wireless Propagation Letters 15, 1681-1684 (2016).
14. V. Bacot, M. Labousse, A. Eddi, M. Fink, and E. Fort, "Time reversal and holography with spacetime transformations," Nature Physics 12, 972-977 (2016).
15. A. Akbarzadeh, N. Chamanara, and C. Caloz, "Inverse prism based on temporal discontinuity and spatial dispersion," Optics letters 43, 3297-3300 (2018).
16. C. Caloz and Z.-L. Deck-Léger, "Spacetime metamaterials—part I: general concepts," IEEE Transactions on Antennas and Propagation 68, 1569-1582 (2019).
17. Z. Yu and S. Fan, "Complete optical isolation created by indirect interband photonic transitions," Nature photonics 3, 91-94 (2009).
18. X. Guo, Y. Ding, Y. Duan, and X. Ni, "Nonreciprocal metasurface with space–time phase modulation," Light: Science & Applications 8, 1-9 (2019).
19. A. E. Cardin, S. R. Silva, S. R. Vardeny, W. J. Padilla, A. Saxena, A. J. Taylor, W. J. Kort-Kamp, H.-T. Chen, D. A. Dalvit, and A. K. Azad, "Surface-wave-assisted nonreciprocity in spatio-temporally modulated metasurfaces," Nature communications 11, 1-9 (2020).
20. V. Pacheco-Peña and N. Engheta, "Temporal aiming," Light: Science & Applications 9, 1-12 (2020).
21. A. M. Shaltout, J. Fang, A. V. Kildishev, and V. M. Shalaev, "Photonic time-crystals and momentum band-gaps," in CLEO: QELS_Fundamental Science, (Optical Society of America, 2016), FM1D. 4.
22. J. S. Martínez-Romero, O. Becerra-Fuentes, and P. Halevi, "Temporal photonic crystals with modulations of both permittivity and permeability," Physical Review A 93, 063813 (2016).
23. Z. Hu, N. He, Y. Sun, Y. Jin, and S. He, "Wideband high-reflection chiral dielectric metasurface," Prog. Electromagn. Res 172, 51-60 (2021).
24. D. Ramaccia, A. Alù, A. Toscano, and F. Bilotti, "Temporal multilayer structures for designing higher-order transfer functions using time-varying metamaterials," Applied Physics Letters 118, 101901 (2021).
25. F. R. Morgenthaler, "Velocity modulation of electromagnetic waves," IRE Transactions on microwave theory and techniques 6, 167-172 (1958).
26. C. Caloz and Z.-L. Deck-Léger, "Spacetime metamaterials—Part II: Theory and applications," IEEE Transactions on Antennas and Propagation 68, 1583-1598 (2019).
27. K. Qu, Q. Jia, M. R. Edwards, and N. J. Fisch, "Theory of electromagnetic wave frequency upconversion in dynamic media," Physical Review E 98, 023202 (2018).
28. Y. Xiao, D. N. Maywar, and G. P. Agrawal, "Reflection and transmission of electromagnetic waves at a temporal boundary," Optics Letters 39, 574-577 (2014).